\documentclass[10 pt]{article}
\usepackage{graphicx}
\usepackage{dcolumn}
\usepackage{epstopdf}
\usepackage{bm}
\usepackage{graphicx}
\usepackage{subfigure}
\usepackage{color}
\usepackage{latexsym}
\usepackage{amsfonts}

\usepackage{amssymb,amsfonts,amsmath}

   \def \S{\Sigma}

\def \>{\rangle} 
\def \<{\langle} 
 
\def\w{\omega}
\def\be{\begin{equation}} 
\def\ee{\end{equation}} 
\newcommand \bea {\begin{eqnarray} } 
\newcommand \eea {\end{eqnarray}}

\begin{document}

\title{A quantitative comparison of
 sRNA-based and protein-based  gene regulation}

\author{Pankaj Mehta$^{1,\star}$, Sidhartha Goyal$^{2,\star}$ and Ned S. Wingreen$^{1}$ \\
$^1$Department of Molecular Biology, Princeton University, Princeton, NJ 08544 \\ and
$^2$Department of Physics, Princeton University, Princeton, NJ 08544 \\ \\
$^{\star}$ Corresponding authors. P Mehta, Department of Molecular Biology, \\ Princeton University, Princeton, NJ 08544, USA. Tel: 609 258 8696; \\ Fax: 1 609 258 8616; Email: pmehta@princeton.edu and \\S Goyal, Department of Physics, Princeton University, \\ Princeton, NJ 08544, USA. Tel: 1 609 240 9316;  \\ Fax: 1 609 258 8616; Email: goyal@princeton.edu}

\maketitle

$\\$
Subject categories: Metabolic and regulatory networks; Signal Transduction; RNA

$\\$

$\\$
Character count: 47450

\newpage
\section*{Abstract}
    
Small, non-coding RNAs (sRNAs) play important roles as genetic regulators in prokaryotes. sRNAs act post-transcriptionally via complementary pairing with target mRNAs to regulate protein expression. We use a quantitative approach to compare and contrast sRNAs with conventional  transcription factors (TFs) to better understand the advantages of each form of regulation. In particular, we calculate the  steady-state behavior, noise properties, frequency-dependent gain (amplification), and dynamical response to large input signals of both forms of regulation. While the mean steady-state behavior of sRNA-regulated proteins exhibits a distinctive tunable threshold-linear behavior, our analysis shows that transcriptional bursting leads to significantly higher intrinsic noise in sRNA-based regulation than in  TF-based regulation in a large range of expression levels and limits the ability of sRNAs to perform quantitative signaling. Nonetheless, we find that sRNAs are better than TFs at filtering noise in input signals. Additionally, we find that sRNAs allow cells  to respond rapidly to large changes in input signals. These features suggest a ``niche" for sRNAs in allowing cells to transition quickly yet reliably between distinct states. This functional niche is consistent with the widespread appearance of sRNAs in stress-response and quasi-developmental networks in prokaryotes.  ({\bf Supporting Information available at www.princeton.edu/$\sim$pmehta/publication.html})

\newpage

\section*{Introduction}

It is now clear that small non-coding RNAs (sRNAs) 
play a crucial role in prokaryotic gene regulation as both positive and negative regulators. sRNAs are involved in many biological functions including quorum sensing ({Lenz {\it et al}, 2004; Fuqua {\it et al}, 2001}), stress response and virulence factor regulation ({Gottesman, 2004}; Majdalani {\it et al}, 2005; Storz {\it et al}, 2004; Storz {\it et al}, 2005), and the regulation of outer-membrane proteins 
({Vogel and Papenfort, 2006; Guillier {\it et al}, 2006}). One major class of prokaryotic sRNAs (antisense sRNAs) negatively regulate proteins by destabilizing the target protein's mRNA (Fig.~\ref{fig:figure1}).  These $\sim$ 100 bp antisense sRNAs prevent translation by binding to the target mRNAs in a process mediated by the RNA chaperone Hfq ({Lenz {\it et al}, 2004; Gottesman, 2004}). Upon binding, both the mRNAs and sRNAs are degraded ({Gottesman, 2004}), suggesting that prokaryotic sRNAs
- unlike their eukaryotic counterparts  - act stoichiometrically on their targets.  Other antisense sRNAs positively regulate protein expression by  promoting ribosome binding to target mRNAs, also in a stoichiometric
fashion ({Gottesman, 2004}).

Whereas transcription factor (TF)-based regulation is ubiquitous in
prokaryotic gene circuits ({Ptashne and Gann, 2001}), thus far
sRNAs have largely been found in circuits responding to strong environmental cues (e.g. extreme nutrient limitation). This leads to a natural question: are transcriptional regulation by TFs and post-transcriptional regulation by sRNAs  distinctly well suited for different biological tasks? 

To address this question, we report a quantitative comparison of the signaling properties of TF-based and sRNA-based gene regulation. In general, a signaling system can be characterized by how it processes different types of inputs.  We  therefore treat TF-based and sRNA-based regulation as signal-processing systems with an input signal -- the average concentration of the TFs controlling RNA transcription rates -- and an output signal -- the average level of the regulated protein ({Ptashne and Gann, 2001}) --  and calculate engineering properties of the system such as the steady-state behavior, noise properties, frequency-dependent gain (amplification), and dynamical response to large input signals ({Detwiler {\it et al}, 2000})  (see Fig.~\ref{fig:figure1}.)



\section{Results}

In the main text, we focus on the case where sRNAs negatively regulate a target mRNA. Positive regulation by  sRNAs is discussed  in the Supplementary Information (SI). 
Post-transcriptional regulation via sRNAs is modeled using mass-action 
equations with three molecular species: the number of sRNA molecules $s$, 
the number of target mRNA molecules $m$, and 
the number of regulated protein molecules $p$ ({Elf {\it et al}, 1995; Elf {\it et al}, 2003};  Lenz {\it et al}, 2004; Levine {\it et al}, 2007;  Mitarai {\it et al}, 2007; Shimoni {\it et al}, 2007). The effect of intrinsic noise is modeled by Langevin terms, $\hat{\eta}_j$, 
that describe the statistical fluctuations in the underlying
biochemical reactions ({van Kampen, 1981}). The kinetics of the various species are 
described by the differential equations
\bea
\frac{ds}{dt} &=& \alpha_s -\tau_{s}^{-1}s - \mu ms  + \hat{\eta}_s + \hat{\eta}_\mu   \nonumber \\
\frac{dm}{dt} &=& \alpha_m - \tau_{m}^{-1} m - \mu ms + \hat{\eta}_m + \hat{\eta}_\mu \nonumber \\
\frac{dp}{dt} &=& \alpha_p m - \tau_p^{-1} p + \hat{\eta}_p \,. 
\label{sRNAdeq}
\eea
The terms can be interpreted as follows. 
sRNAs (mRNAs) are transcribed at a rate
$\alpha_s$ ($\alpha_m$), and are degraded  
at a rate $\tau_{s}^{-1}$ ($\tau_{m}^{-1}$). Additionally, 
sRNAs and mRNAs are {\it both} stoichiometrically degraded by pairing
via Hfq at a rate 
that depends on the sRNA-mRNA interaction strength $\mu$ . Proteins 
are translated from mRNAs at a rate
$\alpha_p$ and are degraded at a rate $\tau_p^{-1}$.

The Langevin terms, $\hat{\eta}_j$, model intrinsic noise by treating
the birth and death processes of the various species in Eq.~(\ref{sRNAdeq}) as independent Poisson processes ({van Kampen, 1981}). 
$\hat{\eta}_s$, $\hat{\eta}_m$, and $\hat{\eta}_p$ model the noise in
the creation and degradation of individual sRNAs, mRNAs, 
and the regulated protein, respectively. $\hat{\eta}_\mu$ models 
sRNA-mRNA mutual-degradation noise.  The Langevin terms are
characterized within the linear-noise approximation by two-point
time-correlation functions  ($j=s, m, p, \mu$) , which for steady states take the form
\be
\<\hat{\eta}_j(t)\hat{\eta}_j(t^\prime)\>= \sigma_j^2 \delta(t-t^\prime) \mbox{\hspace{1in}} 
\label{nocrosscorr}
\ee
with  $\sigma_s^2 =\alpha_s + \tau_s^{-1} \bar{s} $,  $\sigma_m^2 =\alpha_m + \tau_k^{-1} \bar{m} $,  $\sigma_p^2 = 2\tau_p^{-1} \bar{p} $, and $\sigma_{\mu}^2=\mu\bar{m}\bar{s}$ where $\bar{s}$, $\bar{m}$ and $\bar{p}$ denote the mean number  of sRNA, mRNA, and protein molecules, respectively. Notice we have separated the noise due to RNA production and degradation, $\hat{\eta}_s$ and $\hat{\eta}_m$, from the noise due to the binary reaction between mRNAs and sRNAs, $\hat{eta}_\mu$. This allows us to write Eq. \ref{nocrosscorr} in terms of four independent Langevin terms while still capturing the cross-correlation between sRNAs and mRNAs.

Recent evidence suggests that prokaryotic 
transcription may occur with RNA molecules being made in
short intense bursts ({Golding {\it et al}, 2005}). The effects of
transcriptional bursting can be incorporated into our model by allowing two states of gene activation, as reviewed below  (for a detailed discussion see {Paulsson, 2005}). Specifically, genes can be in a transcriptionally inactive ``off" state or  in a transcriptionally active ``on" state. 
The {\it average} transcription rate of 
RNA,  $\alpha_j$ ($j=m,s$) in Eq.~(\ref{sRNAdeq}),
is then related to the probability of the relevant
gene being on, $g^{\rm on}_j$, by 
\be 
\alpha_j = g^{\rm on}_j\alpha^{\rm on}_j
\label{transavgrate}
\ee
with $\alpha^{\rm on}_j$ being the mean transcription rate of the relevant RNA when the gene is always on. We model the dynamics of a repressor-controlled gene using the equation
\be
\frac{dg^{\rm on}}{dt}= k_-(1-g^{\rm on}) - k_+g^{\rm on} + \hat{\eta}_g,
\label{genedynamics}
\ee
where $k_-$ and $k_+$ are the unbinding and binding rates of the repressor and
$\hat{\eta}_g$ is a Langevin noise term. At steady state, it follows from the fluctuation-dissipation theorem that $\<\hat{\eta}_g(t)
\hat{\eta}_g(t^\prime) \> =\sigma_g^2 \delta(t-t^\prime)$ with
$\sigma_g^2= 2k_+ g^{\rm on}$ ({Bialek and Setayeshgar, 2005}). Thus, a full model that
includes transcriptional bursting is described by Eq.~\ref{sRNAdeq} in
conjunction with Eqs.~(\ref{transavgrate}) and (\ref{genedynamics}).

 For completeness, we also briefly review the equations describing transcriptional regulation 
 ({Elowitz {\it et al}, 2002}; Thattai and van Oudenaarden, 2001; Paulsson, 2004;  {Swain {\it et al}, 2002}).  The kinetics of  transcription regulation is  modeled using the Langevin equations
\bea
\frac{dm}{dt} = \alpha_m - \tau_{m}^{-1}m + \hat{\eta}_m \nonumber \\
\frac{dp}{dt} = \alpha_p m - \tau_p^{-1}p  + \hat{\eta}_p
\label{simplegeneexpression}
\eea
with $m$  the  number of mRNA molecules, $p$ the number of 
proteins, $\alpha_m$  the
average rate of transcription, $\alpha_p$ the average
rate of translation,  and $\tau_{m}^{-1}$ and
$\tau_p^{-1}$ the first-order degradation rates 
of mRNA molecules and proteins, respectively.  The two Langevin terms, $\hat{\eta}_m$ and $\hat{\eta}_s$, model noise in the synthesis and degradation of the mRNA and protein, respectively,  (see SI) and obey  the equations ( $j=m, p$)
\be
\<{\hat{\eta}}_{j}(t) \hat{\eta}_{j}(t^\prime)\> 
= 2 \tau_{j}^{-1} \bar{j} \delta(t-t^\prime).
\label{Pnoise1}
\ee
The effects of transcriptional bursting can also be included in this model using Eqs. (\ref{transavgrate}) and (\ref{genedynamics}).


\subsection*{Mean steady-state protein number }
  
The mean steady-state protein number for regulation via sRNAs can be approximated
by ignoring the Langevin terms and setting the 
time-derivatives to zero in Eq.~\ref{sRNAdeq} 
 (see Supporting information (SI) and {Paulsson, 2004, Levine {\it et al}, 2007}) The mean as calculated within this mean-field approximation may differ from the actual mean especially  where noise is large. Nonetheless, the qualitative  steady-state behavior of the mean can  be understood within this approximation.
 
As shown in ({Levine {\it et al}, 2007}) and ({Elf {\it et al}, 2005}), the mean protein number exhibits a threshold-linear
behavior as a function of the mRNA transcription rate $\alpha_m$, with the threshold at $\alpha_s$ 
(see Fig.~\ref{fig:steadystate}). This behavior should be 
contrasted with transcriptional regulation via TFs for which the mean protein number is a linear
function of $\alpha_m$ ~({Elowitz {\it et al}, 2002}; Thattai and van Oudenaarden, 2001; Paulsson, 2004;  {Swain {\it et al}, 2002}).
For sRNA-based regulation, the mean steady-state protein number depends on RNA transcription
rates only through the difference $\alpha_m -\alpha_s$, and 
this dependence can be characterized by 
three distinct regimes: repressed $\alpha_s \gg \alpha_m$, expressing 
$\alpha_s \ll \alpha_m$, and a crossover regime $\alpha_s \approx
\alpha_m$. Increasing the sRNA-mRNA interaction strength $\mu$ results in a sharper crossover between the repressed and expressing regimes. The dashed line in Fig.~\ref{fig:steadystate} depicts the $\mu \rightarrow \infty$
threshold-linear behavior.

In the repressed regime, on average, there are many more sRNAs transcribed than 
mRNAs. Consequently, almost all free mRNAs are quickly bound by sRNAs and
degraded. This results in low levels of expression of the
regulated protein. By contrast, in the expressing regime the average number of mRNAs greatly exceeds the 
number of sRNAs. The sRNAs  
degrade only a small fraction of the total mRNA population so mRNAs accumulate and are translated into proteins. 

\subsection*{Signal transduction}

To compare the signal-transduction properties of sRNA-based regulation with TF-based
regulation, we consider the two regulation schemes as signal-processing systems. Figure~\ref{fig:figure1} depicts how sRNA-based regulation, e.g. in quorum sensing, can be viewed as a signal-processing system (see also SI and Fig. \ref{SI1}). In the context of quorum sensing,  the input signal is the time-averaged  number 
of phophorylated  LuxO (LuxO$\sim$P) molecules in the cell, which, after a series of intermediate biochemical reactions, 
is converted into the output signal, the average number of LuxR
molecules. Fluctuations in LuxO$\sim$P and LuxR 
about their averages can be thought of as the input and output noise, respectively. 
The noise in the output is a combination of  {\it input noise} -- fluctuations in the input signal, {\it intrinsic noise} -- stochasticity inherent in gene regulation, and {\it extrinsic noise} -- other sources of noise impinging 
on the signal-processing system not explicitly considered in the model, such as ribosome and RNA-polymerase fluctuations.  

The fidelity of a signaling system is ultimately limited by the output noise of the system.  The output noise,  defined as  the ratio of the variance in the output-protein  number to the square of the mean output-protein number, can be thought of as the square of the  ``percentage error'' in the output. The higher the output noise, the poorer the signaling fidelity of a
gene-regulation scheme. Thus, examining the noise properties of sRNA-based and transcription-factor gene regulation is important for comparing these two forms of gene regulation.

 Gene regulation takes place as part of a larger genetic and biomolecular network whose purpose is to convert a measured signal into a concentration of the regulated protein. A simple but important observation is that sRNA-based regulation also requires protein regulators in order to couple to external signals. In particular, a protein regulator is necessary to vary the transcription rate of the sRNAs in response to an input.  For this reason, we take as the input signal to both systems a protein  that either transcriptionally regulates the relevant protein directly  or else transcriptionally regulates the sRNAs. In the case of direct transcriptional regulation, the protein regulator acts as a repressor whereas for post-transcriptional, sRNA- based regulation, it acts as an activator (see Fig. \ref{SI1}).  Furthermore, the kinetics of the protein regulator are chosen to be identical in both cases. The upstream components of the network that controls the level of the relevant protein regulator  are also assumed to be identical. This allows for a principled comparison of the two regulatory schemes.

\subsubsection*{Intrinsic noise}

Gene regulation is intrinsically noisy. In this paper, we define intrinsic noise as the fluctuations in the output-protein number, given a fixed steady-state input, due to the stochastic nature of the underlying biochemical reactions. When calculating intinsic noise we neglect the contributions to output noise from fluctuations in the input and from extrinsic noise sources such as variations in the number of ribosomes and RNA polymerase molecules (see Fig. \ref{fig:figure1}). 


 We start by summarizing the noise properties of transcriptional regulation. For ordinary transcriptional regulation by a repressor, the
intrinsic noise -- defined 
as the variance in protein number divided
by the mean protein number squared,  $\sigma_p^2/\bar{p}^2$, is given by
({Elowitz {\it et al}, 2002}; Thattai and van Oudenaarden, 2001; Paulsson, 2004; {Swain {\it et al}, 2002}; Golding {\it et al}, 2005; Paulsson, 2005) (and  SI):
\be
\eta_{\rm int}^{\rm TF}= \frac{\sigma_p^2}{\bar{p}^2} = \frac{(1+b)}{\bar{p}} +
\frac{(p_{\mathrm{max}}-\bar{p})\bar{p}^2}
{\bar{p}+k_-\tau_p
  p_{\mathrm{max}}}
\label{noiseTF}
\ee  
where $ b= \alpha_p \tau_m$
is the protein burst size (the average number of proteins made from an mRNA molecule) and
$p_{\mathrm{max}}=\alpha_{m}\alpha_p \tau_m\tau_p$ is the mean protein level
in the absence of repressor. The first term in Eq.~(\ref{noiseTF}) captures
the noise due to translational bursting (the protein burst from each mRNA due to the  translation of multiple proteins from each mRNA molecule) and the second captures the noise due to transcriptional bursting (the RNA burst while no repressor is bound). The transcriptional
bursting contribution is typically much smaller than that of
translational bursting  since   the unbinding rate of the repressor is generally  much faster
than the protein degradation rate, $k_-\tau_p \gg  1$ . Consequently, the intrinsic noise for protein-based regulation is often approximated as $\sigma_p^2/\bar{p}^2 \approx (1+b)/\bar{p}$.

The intrinsic noise of an sRNA-regulated protein
differs significantly from that of a transcriptionally regulated protein. Noise in stoichiometrically coupled systems such as sRNA-based gene regulation has been studied previously ({Elf and Ehrenberg, 2003; Elf {\it et al}, 2003; Paulsson and Ehrenberg, 2001}). It was found by Elf {\it et al, 2005} that the ultrasensitivity of stoichiometric systems in the crossover regime necessarily gives rise  to enhanced stochastic fluctuations.  This "near-critical" behavior was related to the behavior at phase transitions where fluctuations also diverge ({McNeil and Wall, 1974}). We have extended these previous analyses to the context of gene regulation by sRNAs, and have calculated the intrinsic protein noise  within the linear-noise approximation (see SI and ({van Kampen, 1981; Elf and Ehrenberg, 2003})), including the effects of transcriptional and translational bursting. We have checked our results using exact stochastic simulations (see SI and Figs. SI-2 and SI-3). The simulations confirm the existence of three regimes and verify  that noise is enhanced in the crossover region due to critical fluctuations.

The full expressions for the intrinsic noise are lengthy and in the main text we present only our major findings.  Figs. \ref{fig:noiseburst} and \ref{fig:fig4} show typical intrinsic noise profiles as functions of the transcription-rate ratio, $\alpha_s/\alpha_m$, and of the average protein level of the regulated protein, for various magnitudes of transcriptional
bursting. For a given sRNA-mRNA interaction strength $\mu$,
the intrinsic noise increases with larger transcriptional bursts (smaller $k_-$). Furthermore, for a fixed $k_-$, the intrinsic noise increases with increasing sRNA-mRNA interaction strength $\mu$ (see  Fig. SI-1 and {Elf and Ehrenberg, 2003}). The intrinsic noise is small in the repressed regime $\alpha_s \gg \alpha_m$,
and shows a pronounced peak in the crossover region, $\alpha_s \approx \alpha_m$ 
 (see Fig. \ref{fig:noiseburst}) as expected for a stoichiometric system. We have also obtained simplified, asymptotic expressions for the noise in the repressing and expressing regimes  when $\tau_m \ll \tau_p$, and there is no transcriptional bursting (see SI). The expressions for the intrinsic noise in the repressing and expressing regimes are 
given by, respectively:
\be
\eta_{\rm int}^{\rm rep} \approx \frac{(1+ b_{\rm eff})}{\bar{p}}
+ \frac{\bar{p}}{p_{\rm max}} \frac{\mu  \tau_{s} \tau_{m}}{\tau_p}, 
\label{noiseRR}
\ee
(where  $p_{\rm max}=\alpha_p\tau_m \alpha_m \tau_p$ and $ b_{\rm eff}= b(\bar{p}/p_{\rm max}) \ll b$  is the new ``effective" protein
burst size (see SI  and ({Levine {\it et al}, 2007})),
and
\be
\eta_{\rm int}^{\rm exp} \approx \frac{(1+b)}{\bar{p}}+
\frac{p_{\rm max}}{\bar{p}^4}\frac{\alpha_p^3 \tau_p}{\mu ^2} .
\label{noiseER}
\ee
We have written these expressions so that the contribution
of sRNA-mRNA mutual-degradation noise is  contained entirely in the second term of Eqs.( \ref{noiseRR}) and( \ref{noiseER}).

 Comparing the intrinsic noise of protein- and sRNA- based regulators in Fig. \ref{fig:fig4}, we observe that sRNA regulators are significantly less noisy than TFs in the repressed regime.  The dominant source of intrinsic noise for a TF-regulated protein, in the limit $\tau_m \ll \tau_p$, is that proteins are made in bursts of average size $b \gg 1$. For an sRNA-regulated protein, the average size of a protein burst, $b_{\rm eff}$,  is much smaller (see Eq. \ref{noiseRR}). This can be understood by noting that  there are many more sRNAs than mRNAs in the repressed regime, and therefore any free mRNA is quickly bound by an sRNA and degraded. This leads to a reduction in the effective mRNA lifetime and consequently a reduced $b_{\rm eff}$ ({Levine {\it et al}, 2007}).  The reduction in effective mRNA lifetimes and  intrinsic noise takes place even when mRNAs and sRNAs are produced in bursts.

 The fidelity of a signaling system can be characterized by the output noise $(\sigma_p^{\rm total})^2/\bar{p}^2$.  In general, high-fidelity signaling requires $(\sigma_p^{\rm total})^2/{\bar p}^2 \ll
1$.  Thus, from Fig.~\ref{fig:fig4} it is clear
that over a large range of output protein levels the large intrinsic noise due to transcriptional bursting makes it difficult for sRNAs to perform high-fidelity signaling.

 One of the most striking features of  Fig. \ref{fig:fig4} is that sRNA-based regulation is much more sensitive to transcriptional bursting than is protein-based regulation.  For sRNAs, transcriptional bursting  greatly enhances the near-critical fluctuations because the production of RNAs in bursts  increases the anti-correlated sRNA-mRNA fluctuations in
 the crossover regime (see {Elf {\it et al}, 2003; Elf and Ehrenberg, 2003 for more on the near-critical fluctuations).  In contrast for transcriptional regulation directly by a TF, the contribution of transcriptional bursting to the intrinsic noise is relatively small  for most choices of parameters ( see Fig. \ref{fig:fig4}). Since recent experiments suggests that prokaryotic  transcription may generically produce RNAs in bursts ({Golding {\it et al}, 2005}), this is likely to be a physiologically relevant effect for sRNA-based gene regulation.

 The large intrinsic noise in the crossover regime, $\alpha_s \approx \alpha_m $ can be understood by
considering the special case  $\alpha_s=\alpha_m$ for very strong sRNA-mRNA binding,
$\mu \rightarrow \infty$.  In this limit,  sRNAs and mRNAs, transcribed at 
the same average rate,  quickly bind to each other and degrade 
and almost no protein is made. However, once in a
while there is a fluctuation that produces more mRNAs than
average. In this case, unless there is a corresponding fluctuation in
sRNAs, the mRNAs cannot be degraded by sRNA-mRNA binding.
The  mRNAs produced in such a fluctuation will  degrade by the
usual slow degradation rate $\tau_m^{-1}$ resulting in a large burst of protein production, contributing to the large intrinsic noise. Transcriptional bursting further increases the magnitude of the aforementioned sRNA and mRNA fluctuations and consequently further increases the intrinsic noise in the crossover regime.

\subsubsection*{Gain and Filtering}
We now consider, in the absence of noise, the change in output-protein number about some
steady state or ``operating point'' in response to a small, time-varying input signal. A small time-varying change from the
steady-state value of the number of proteins controlling the sRNA transcription rate,  
$\delta c(t)= c(t) -\bar{c}$, results in a corresponding time-varying
change of the output-protein number from its steady-state value, $\delta
p(t)= p(t) - \bar{p}$.  For small enough
signals, the dynamics are captured by linearized versions
of the   mass-action equations (Eq.~\ref{sRNAdeq})
(see SI).  In the frequency-domain, the relationship
between the output protein response at frequency $\omega$ and
the  input signal at frequency $\omega$ takes the simple form
\be
\delta \tilde{p}(\omega)=\tilde{g}(\w)\delta\tilde{c}(\w)
\ee
where the frequency-dependent gain  is given by
\be
\tilde{g}(\w)\propto \underbrace{\frac{k_+ g^{\rm on}}{i\w+\tau_g^{-1}}}_{\mathbf{I}}
\underbrace{ \frac{\mu\bar{m}\alpha_s^{\rm on}}{(i\w+\tau_+^{-1})(i\w+\tau_-^{-1})}}_{\mathbf{II}}
 \underbrace{\frac{\alpha_p}{i\omega + \tau_p^{-1}}}_{\mathbf{III}},
 \ee
 with $\tau_g=k_-+k_+$ the characteristic time the sRNA gene is "on" and $\tau_\pm$ two times related to -- and of the same order of magnitude as-- the mRNA and sRNA lifetimes (see SI for exact definition of $\tau_\pm$).
Each term of the form $(i\omega + \tau^{-1})^{-1}$ can be interpreted as
a low-pass filter with cut-off frequency $\tau^{-1}$. The four low-pass filters in the frequency-dependent gain come from  different intermediate-steps: $\mathbf{I}$ from the binding-unbinding of the protein regulator (activator), $\mathbf{II}$ from the transcription of RNAs and 
the sRNA-mRNA interaction, and $\mathbf{III}$ from the translation 
of mRNAs into proteins.
The amplitude of the frequency-dependent gain decreases rapidly $\propto\omega^{-4}$ at high frequencies.
This can be compared to the the gain in TF-based regulation 
which has only three low-pass filters and  falls of at high frequencies $\propto\omega^{-3}$ (see Figure \ref{fig:fig5} and SI) .  Thus, we conclude that sRNA-based regulation is less
sensitive to high-frequency input noise than is TF-based 
regulation.

The underlying reason for the enhanced noise filtering properties of sRNAs is that sRNA-based 
regulation involves an additional step when compared to transcriptional regulation. Namely, the input signal from upstream components in the genetic network is transmitted to the mRNAs encoding the output protein  via  sRNAs, which corresponds to an additional noise filter.  This extra filtering could also be achieved by introducing an additional layer of transcriptional regulation in the genetic network. However, adding an extra layer of  transcriptional regulation also leads to a slower kinetic response of the signaling network to changes in  the input signal  because an additional protein regulator must be synthesized or degraded in order to transmit signals.  This kinetic cost is much smaller for sRNA-based regulation (see below). Consequently, sRNA-based regulation allows for an extra layer of noise filtering without sacrificing the ability to respond quickly to changes in input.

The above results hold only when the input signal is coupled to the sRNAs. Small input signals can also  modulate the transcription of the protein-coding  mRNAs instead of the 
sRNAs. In this case, at  high frequencies,
the gain falls off as $\propto \omega^{-3}$ as in TF-based regulation since the input signal does not pass through the sRNAs (see SI). Thus, coupling the input signal to sRNAs instead of
mRNAs is necessary to achieve the advantageous high-frequency filtering properties of
sRNA-based gene regulation.  This may explain why input signals are often found coupled to the sRNAs rather than to the mRNAs in sRNA-based regulatory circuits.

\subsubsection*{Fidelity of small-signal response}

Intrinsic noise limits the ability of a signaling system to faithfully respond to small signals.
Typically, the ability of a system to transduce small signals is quantified by its gain (amplification factor) ({Detwiler {\it et al}, 2000; Elf {\it et al}, 2003; Elf and Ehrenberg, 2003}). A large gain is interpreted to mean the system can differentiate small changes in the input signal. However, even if the gain is large, if there is also high intrinsic noise -- as is the case in sRNA-based regulation -- it may be impossible to distinguish the output signal from the output noise (see {Detwiler {\it et al}, 2000} and SI).   Furthermore, the gain often depends on how input and output signals are defined (e.g. logarithmic gain vesus linear gain). For this reason,  we consider an alternative  measure to compare the small-signal responses of sRNA- and protein-based regulators, namely the minimal signal that can be faithfully transmitted by the system ({Detwiler {\it et al}, 2000}).

As discussed above, the noise in the output protein limits the detection of small input
signals.  In order for an input signal to be detectable, the
corresponding output signal must be greater than the output noise 
({Detwiler {\it et al}, 2000}). In particular, the power of the output signal must be greater than the power of the output noise. Consider a periodic input signal at a frequency $\omega_0$ and amplitude $\delta c_{\omega_0}$, $\delta c_{\omega_0} e^{i\w_0 t}$.  For
small input signals, the output signal is related to the input signal
by the frequency-dependent gain $g(\omega)$. Thus, the output signal is $O(t) = g(\w_0)\delta c_{\omega_0} e^{i\w_0 t}$ and the power of the output signal is by definition
\bea
{\rm Power}_{\rm sig} &=& \w_0\int_0^{1/\w_0} dt |O(t)|^2 \nonumber \\
&&=|g(\omega_0)\delta c_{\omega_0}|^2. 
\eea
On the other hand, the power of the output noise is calculated by integrating fluctuations over all frequencies, and is given within the linear-noise approximation by the expression
\be
{\rm Power}_{\rm noise} =  \int \frac{d\omega^\prime}{2\pi} \frac{d\omega^{\prime \prime}}{2\pi}
\delta p^*(\omega^\prime)\delta p(\omega^{\prime \prime })=\sigma_p^2
\ee 
 where $\delta p(\omega^\prime)$ is just the fluctuation in the output-protein level at a frequency $\omega$ due to intrinsic noise as calculated in the SI. For a signal to be detectable, we must have
 \bea
 {\rm Power}_{\rm sig} &\ge& { \rm Power}_{\rm noise} \nonumber \\
|g(\omega_0)\delta c_{\omega_0}|^2 &\ge&  \sigma_p^2.
 \label{powerineq}
 \eea

For a step-input signal with amplitude $\delta c_{\circ}$ ($\omega_0 \rightarrow 0$ in the above expressions), the requirement that the output signal is larger than the noise sets a lower-bound on the detectable input signal
 $\delta{c}^{\rm min}_{\circ} \ge  \sigma_p/{g}_{\circ}$ ({Detwiler {\it et al}, 2000}).  Of course, by time-averaging the output, one can reduce the output noise and hence detect smaller signals, but this does not affect our comparison. Therefore, we  computed the minimum input signal without time-averaging for both sRNA-based and TF-based
regulation and found that, for  even moderate amounts of transcriptional bursting, protein regulators are better than sRNAs at responding to small signals across the whole range of output-protein levels. At low protein levels (repressed regime), the minimum detectable signal for sRNA-based regulation is larger due to the lower gain for sRNA-based regulation than for TF-based regulation. At intermediate to high levels of output protein (crossover and expressing regimes), the minimum detectable signal for sRNAs is also larger due to the large protein noise $\sigma_p^2$ arising from transcriptional bursting for sRNA-based regulation. 

 Consequently, contrary to previous speculations (Levine {\it et al}, 2007), results indicate that sRNA-based regulation is unlikely to be useful for amplifying small signals despite the large gain of sRNA-based regulation in the crossover region. Our results also imply that it is more advantageous to use transcription factor-based regulation than sRNA-based regulation in genetic networks designed to respond to small changes in upstream components.

\subsection*{Large-signal response}

In nature, an organism may benefit from  switching quickly between two different
gene-expression states in response to a large persistent input signal. We have compared here the rates at which a regulated protein can switch between an
``off" and ``on" state in response to an input signal when its mRNA is directly regulated by a TF or indirectly regulated by an sRNA.

Fig.~\ref{largesignal} shows the time evolution of the average mRNA level for both sRNA-based and TF-based regulation in response to a step change in the input. The response for sRNA-based regulation depends on the initial conditions, and can be tuned by changing where in the repressed regime the system is initially. In particular, the effective mRNA degradation (and dilution) rate depends on the sRNA pool size and on the sRNA-mRNA interaction strength $\mu$. However, our conclusions do not strongly depend on the choice of parameters (see SI).

We find that using sRNAs to switch protein expression on, i.e., going from low
output-protein number to high output-protein number, is
slower than direct TF regulation. This slower response is due to the
sRNA pool that needs to be depleted before target mRNAs can accumulate.  On the other hand,   sRNA-based regulation can be faster than TF-based
regulation when switching off expression of a protein -- the large input signal rapidly increases the concentration of sRNAs resulting in fast degradation of target mRNAs (see Fig.~\ref{largesignal} and  Shimoni {\it et al}, 2007). The slower response of the sRNA-based regulation at turning on protein expression  stems from the delay introduced by having an  additional  layer of sRNA regulation in the signal transduction pathway when compared with protein-based regulation (see Fig.~\ref{largesignal}). However, this delay is much smaller than that which would be introduced by having an additional layer of transcriptional regulation since the synthesis and degradation rates of proteins are much slower than those of RNAs (see SI for a discussion comparing our results with Shimoni {\it et al}).

  Thus far we have considered the case where a protein is negatively regulated by sRNAs.
 However, a protein can also be positively regulated by sRNAs (see ({Storz {\it et al}, 2004; Hammer and Bassler,  2007} and SI),
and in this case  switching protein expression on   using sRNAs can be faster than TF-based regulation.  Typically, sRNAs positively regulate protein expression by preventing the formation of inhibitory secondary structures that occlude the ribosome binding sites of the regulated mRNA. Since there is generally a background pool of translationally-inactive target mRNAs,   a large input signal that produces sRNAs allows the target mRNAs to be quickly converted into the translationally-active form.

\section*{Discussion}

Small non-coding RNAs (sRNAs) play an important regulatory role in prokaryotic gene circuits. sRNAs are  involved in a variety of critical physiological tasks such as quorum sensing, stress response, and the regulation of outer-membrane proteins. Yet,  sRNAs are not currently thought to be as common as transcription factors  (TFs)  in prokaryotic gene regulatory circuits (at least based on our present knowledge), suggesting sRNAs may be well suited for certain biological tasks but not for others. This paper evaluates the suitability of sRNA-based regulation to particular biological tasks by treating gene regulation as a signal-processing system.

Our analysis shows that for a large (intermediate to high) range of output-protein levels, the intrinsic noise for sRNA-based regulation is much larger than for TF-based regulation. However, even at a high level of transcriptional bursting, sRNA-based regulation is less noisy than TF-based regulation at low protein levels (in the repressed regime) because a large pool of sRNAs shortens the effective mRNA lifetime and buffers against target mRNA fluctuations. Thus, in all cases, protein expression can be kept  off much more reliably by sRNAs than by TFs (see SI for a discussion of the dependence on kinetic parameters). We also find (when the input signal is coupled to the sRNAs) that sRNAs  are better filters of high-frequency input noise than TFs since they implement an extra low-pass filter compared to TFs. 
This is likely to be physiologically relevant since sRNAs are often found in networks that couple to external signals ({Majdalani {\it et al}, 2005}). In such networks, high-frequency noise in the input could arise  from noise in external concentrations or from the fast upstream protein-modification reactions such as phosphorylation-dephosphorylation of a two-component system . sRNAs also allow cells to respond quickly to large changes in  input signal. In particular, sRNAs can quickly turn off negatively regulated genes and quickly turn on positively regulated genes (Shimoni {\it et al}, 2007).  This ability to filter high-frequency noise without comprimising the ability to rapidly respond to input signals is a defining feature of sRNAs. The above characteristics make sRNA-based regulation useful for constructing genetic switches.  In contrast, even for moderate levels of transcriptional bursting,  sRNA-based regulatory circuits  are worse than TFs at transducing small input signals, suggesting  that TFs are likely better suited for quantitative adjustment of protein expression. Additionally, the use of sRNAs in more complex network motifs such as feed-foward loops is likely to give rise to new behaviors (Shimoni {\it et al}, 2007). Our results are summarized in Table \ref{table}.

Indeed, sRNAs are often found in genetic circuits that  switch gene-expression states in response to strong environmental cues.  For example under iron limitation, the sRNA RyhB rapidly shuts off synthesis of several iron-binding proteins, making iron available for essential
proteins ({Masse and Gottesman, 2002}).  In the quorum-sensing network of \emph{Vibrio harveyi} and of the human pathogen \emph{Vibrio cholerae}, multiple sRNAs (Qrr1-5) switch the expression state of the cell in response to external cell density ({Lenz {\it et al}, 2004}). The fast dynamical response of sRNA-based regulation, accelerated by the presence of five sRNAs, may allow the pathogen  \emph{V. cholerae} to quickly switch expression states in response to a  sudden change in the environment -- for example, from a high bacterial cell density in a eukaryotic host  to low cell density in the  marine environment ({Hammer and Bassler, 2007; Zhu {\it et al}, 2002}). Recent modeling work by (Shimoni {\it et al}, 2007) suggests that the kinetic properties of sRNAs  are crucial to understanding the behavior of  {\it E. coli} regulatory circuits involved in responding to osmotic stress.
In both the iron-metabolism and quorum-sensing circuits discussed above, the input signals, iron limitation and cell density, are coupled to the expression of  sRNAs and not to the target mRNAs ({Lenz {\it et al}, 2004; Masse {\it et al}, 2007}), suggesting that filtering input noise may also be an important consideration (see Fig. 4).  

We have considered the case where a single sRNA species regulates a single mRNA species. However, as in the {\it Vibrio} quorum-sensing circuit, multiple sRNAs may regulate multiple mRNAs ({Lenz {\it et al}, 2004; Mitarai {\it et al}, 2007; Repoila {\it et al}, 2003}). Even in such a case, mean steady-state protein numbers are expected to exhibit a threshold-linear behavior with three distinct regimes. The main difference from the single sRNA/mRNA case is that the threshold occurs when the combined sRNA transcription rate exceeds the target mRNA transcription rate ~({Levine {\it et al}, 2007}; Shimoni {\it et al}, 2007).  This may allow sRNAs to prioritize usage of different target mRNAs  ({Levine {\it et al}, 2007}; Mitarai {\it et al}, 2007).

There are additional considerations that may favor sRNA-based or TF-based regulation. For example,   TFs are likely  to be better global regulators than sRNAs -- since sRNAs degrade mRNAs stochiometrically, only a limited number of genes can be regulated by a given sRNA.  Also,  the cost in space on the genome is generally larger for sRNA-based regulation than for direct regulation by TFs  because in the former it is necessary to encode for the sRNA in addition to the regulatory region of the regulatory TF coupling the sRNA to external signals (see Fig. \ref{SI1}) ({Semsey {\it et al}, 2006}).  Aditionally, sRNAs and TFs are likely to respond differently  to  extrinsic noise. For example, one expects  sRNA-based regulation to be less sensitive to global RNA-polymerase fluctuations than TFs since sRNAs and their target mRNAs are affected identically by polymerase abundance ({Paulsson and Ehrenberg, 2001}).  Finally, the metabolic costs of sRNA-based regulation and protein-based regulation may differ (Mitarai {\it et al}, 2007).

In this paper, we have considered gene regulation by non-coding RNAs in prokaryotes. Regulatory RNAs are also found in eukaryotes. In eukaryotes, these regulatory RNAs are believed to act catalytically, not stoichiometrically. Nonetheless, our analysis suggests that, even in eukaryotes, regulatory RNAs are better at keeping protein expression off  than TFs, since in both cases, regulatory RNAs shorten the effective lifetime of their target  mRNAs, thus reducing protein fluctuations. Furthermore, our analysis also suggests that regulatory RNAs in eukaryotes are likely better than TFs at filtering out high-frequency input noise in upstream signals.

 Recently, it has been shown that noise in protein expression may exhibit a universal behavior ({Bar-Even {\it et al}, 2006}). However, our analysis for the intrinsic noise of an sRNA-regulated protein differs significantly from the proposed universal behavior in the presence of transcriptional bursting (see also ~({Tkacik {\it et al}, 2006})). It would be interesting to test our predictions for intrinsic noise experimentally by quantifying intrinsic cell-to-cell variation of a fluorescent protein ({Elowitz {\it et al}, 2002}) alternatively regulated by an sRNA or a TF, particularly with controllable transcriptional bursting ({Blake {\it et al}, 2006}).

The analogy between biochemical circuits and signal-processing systems in engineering provides a general framework for characterizing the signal-transduction pathways found in biology (Detwiler {\it et al}, 2000). Different biological tasks place different requirements on signal-transduction circuits. For example, in chemotaxis, bacteria must respond quickly to changing input signals ({Bialek and Setayeshgar, 2005; Berg, 2003; Keymer, 2006}) whereas in quorum sensing or stress response, reliability may be more crucial than speed. 
One suspects that biological networks exhibit a harmony between network architecture and network function. For this reason, understanding the comparative advantages and disadvantages of different architectures is likely to yield new insights into biological function, as well as new schemes for synthetic circuits.			

\section*{Materials and Methods}

The analyses were carried out using rate-equation models  extended to include stochastic fluctuations and our results were tested using Monte-Carlo (Gillespie) simulations. The equations account for the concentration of each component in the circuit, and for noise around the means of these components. The dynamics of gene regulation was modeled using Langevin equations  for the various species in the system:  mRNAs, sRNAs, and proteins. Using this model, we analyzed the signaling properties of the two regulation schemes, focusing on gain, filtering, and switching times in response to large input signals. For further details see the Supplementary Information.

\section*{Acknowledgments}
We would like to thank  Bonnie Bassler, Matthias Kaschube Anirvan Sengupta, Gasper Tkacik, Chris Waters and Kerwyn C. Huang for helpful discussions and suggestions on the manuscript. This work was partially supported by US National Institutes of Health (NIH) Grant PSO GM071508,  the Defense Advanced Research Projects Agency (DARPA) under grant HR0011-05-1-0057, and the Burroughs Wellcome Fund Graduate Training Program.


\newpage

\section*{Figure legends}

\begin{figure}[h]
\includegraphics[width=0.9\textwidth]{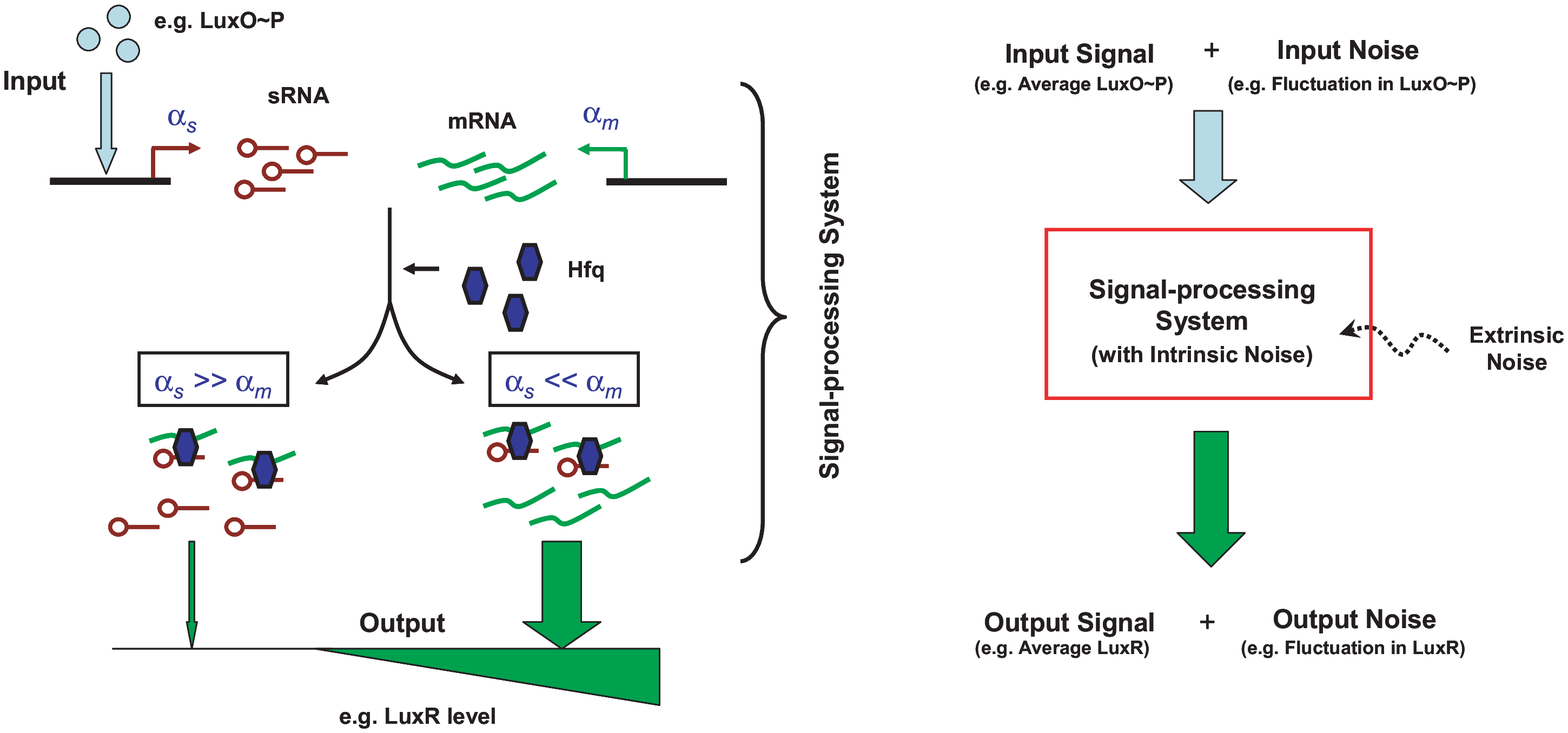}
\caption{Genetic Regulation via sRNAs. Left:
 Small non-coding RNAs (sRNAs) regulate protein expression as part of  a larger 
genetic network with a specific biological task 
(e.g.  quorum sensing in Vibrio bacteria~({Lenz {\it et al}, 2004}). 
The sRNAs (stem loops) regulate 
target proteins by destabilizing target-protein
mRNAs (wavy lines), a stoichiometric process mediated by the 
RNA chaperone Hfq (hexagons). When the rate of sRNA transcription $\alpha_s$
greatly exceeds the rate of mRNA transcription $\alpha_m$, i.e. when $\alpha_s \gg \alpha_m$, 
nearly all the mRNAs are bound by sRNAs and cannot be translated.
By contrast, when $\alpha_m \gg \alpha_s$, there are many more
mRNAs than sRNAs, and protein is highly produced.
Right: The stochasticity (randomness) of cellular processes
results in noise -- statistical fluctuations in the molecular
numbers. It is helpful to classify the total noise in the output ({\it output noise}) into (i) {\it input noise} -- noise in the input signal from upstream-components in the gene circuit,
(ii) {\it intrinsic noise} -- 
noise from stochasticity inherent in gene regulation via
sRNAs, and (iii) {\it extrinsic noise} -- all other sources of noise impinging on the signal-processing system.}
\label{fig:figure1}
\end{figure}

\begin{figure}
\includegraphics[width=0.9\textwidth]{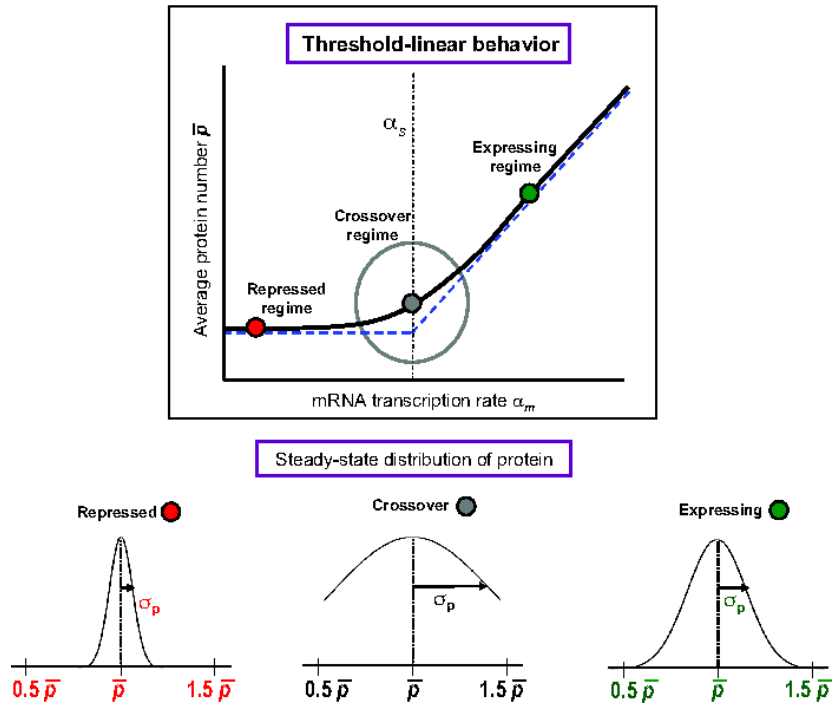}
\caption{Steady-state behavior for gene regulation via
sRNAs. For the regulated protein, the steady-state mean number $\bar{p}$ exhibits
an approximately threshold-linear behavior as a function of the mRNA
transcription rate $\alpha_m$. The threshold is set by
the sRNA transcription rate $\alpha_s$. Protein expression can
be classified into three regimes: repressed ($\alpha_s \gg
\alpha_m$), crossover ($\alpha_s \approx \alpha_m$), and
expressing ($\alpha_s \ll \alpha_m)$. 
In the repressed regime, the average protein number is low. By contrast, the protein number increases almost linearly with $\alpha_m$ in
the expressing regime. The typical behavior of the noise
$\sigma_p$, the standard deviation of the protein number, 
is shown for the three regimes.}
\label{fig:steadystate}
\end{figure}

\begin{figure}
\includegraphics[width=0.9\textwidth]{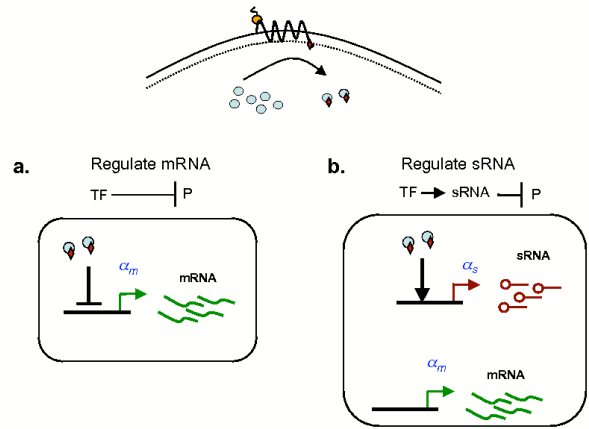}
\caption{Schematic showing our comparison of  transcriptional and post-transcriptional, sRNA-mediated regulation.
We take as the input signal to both systems a protein regulator (blue discs)  that either directly transcriptionally regulates the relevant gene by acting as a repressor or transcriptionally regulates an sRNA acting as an activator. The protein regulator is chosen to have identical kinetic properties in both cases. }
\label{SI1}
\end{figure}

\begin{figure}
\includegraphics[width=0.9\textwidth]{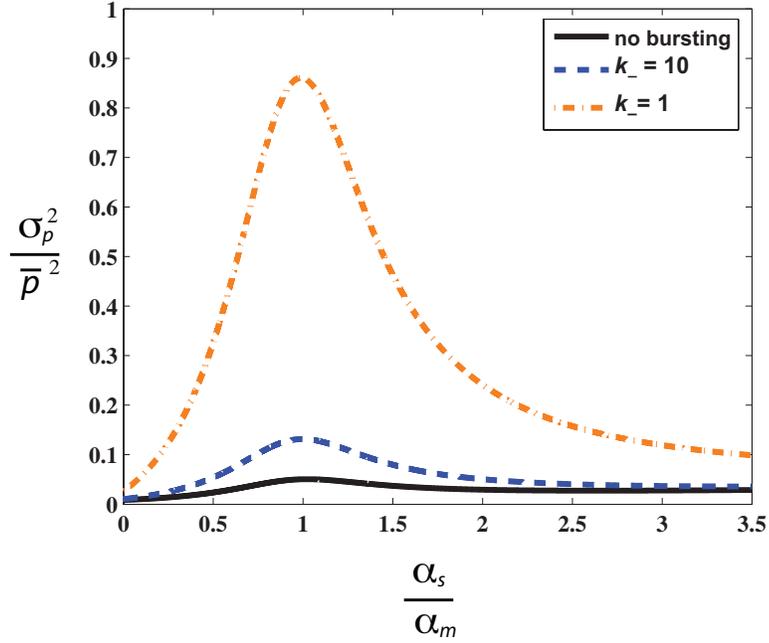}
\caption{Protein noise with or without transcriptional bursting.
Noise in protein expression $\sigma_p^2/\bar{p}^2$
(variance divided by mean squared) as a function of the ratio
of the sRNA and mRNA transcription rates, $\alpha_s/\alpha_m$, for different
levels of transcriptional bursting.  We have assumed 
that both the sRNAs and mRNAs are produced in bursts. The noise peaks in the
crossover regime, $\alpha_s \approx \alpha_m$. A slower unbinding rate $k_-$ for the repressor proteins controlling sRNA  and
mRNA expression results in larger transcriptional bursts.  Parameters are (in
${\text{min}}^{-1}$):
$\alpha_m=3$, $\alpha_m^{\rm on}=10$, $\alpha_{s}^{\rm on}=30$, $\tau_m=10$, $\tau_s=30$,
$\mu =0.02$, $\alpha_p=4 $, $\tau_p=30$, and $k_+$ is adjusted to set the mean protein levels (for a discussion of parameter dependence see Supplementary Information).}
\label{fig:noiseburst}
\end{figure}

\begin{figure}
\includegraphics[width=0.9\textwidth]{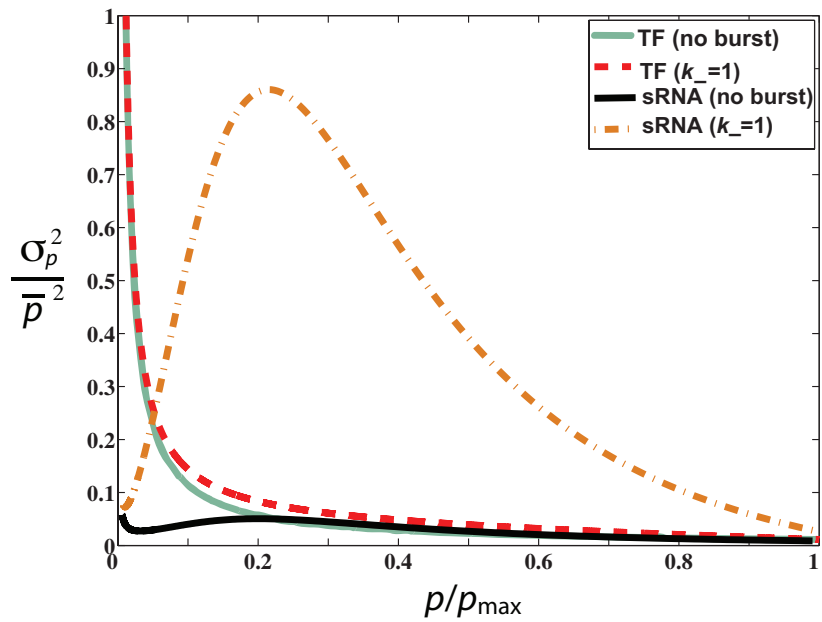}
\caption{Comparison of analytic expressions for the intrinsic protein noise for TF-based and sRNA-based regulation. Shown is the intrinsic noise for sRNA-based regulation as a function of normalized average protein concentration, $\bar{p}/p_{\rm max}$, with and without transcriptional bursting.  All parameters as in Fig.~\ref{fig:noiseburst}.}
\label{fig:fig4}
\end{figure}

\begin{figure}
\includegraphics[width=0.9\textwidth]{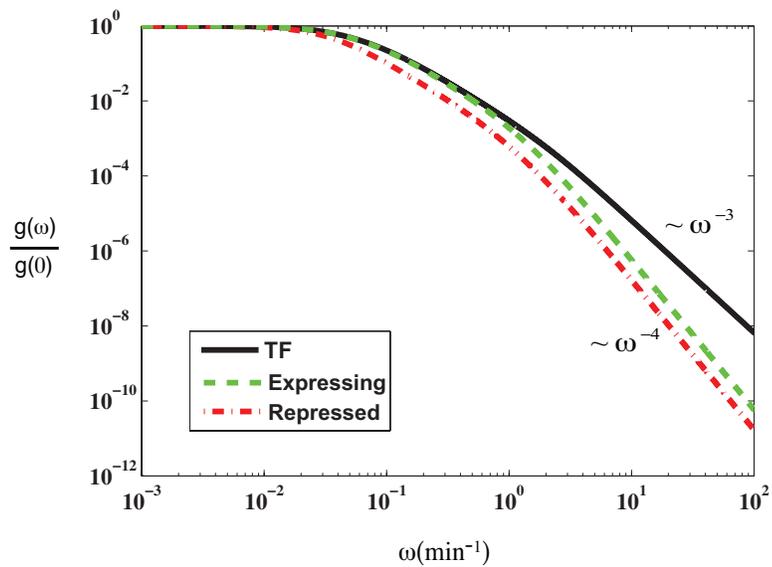}
\caption{Normalized frequency-dependent gain, $g(\omega)/g(0)$, as a function of the frequency, $\omega$,  for a small
input signal  for TF-based regulation and for sRNA-based regulation in the repressed and expressing regimes.   The amplitude of the frequency-dependent gain decreases rapidly $\propto\omega^{-4}$ at high frequencies for sRNAs compared to $\propto \omega^{-3}$ for TFs. Consequently, sRNA-based regulation is less sensitive to high-frequency input noise than is TF-based 
regulation. Parameters as in Fig.~\ref{fig:noiseburst}. }
\label{fig:fig5}
\end{figure}

\begin{figure}
\begin{center}
\includegraphics[width=0.9\textwidth]{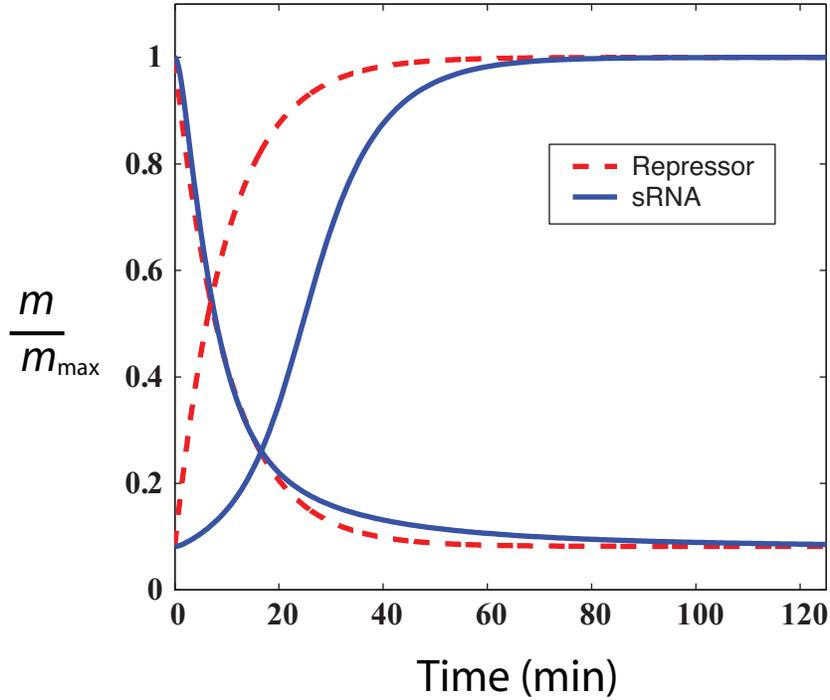}
\end{center}
\caption{Large-signal switching. Normalized mRNA level $m/m_{\rm max}$, as a function of time, in response to step changes in the input, for both the sRNA-based and the TF-based regulation.
Switching from high mRNA level (on state) to low mRNA level (off state) and vice versa. Switching from off to on state has a lag in the case of sRNA-based regulation, whereas the switching time from the on to off state for sRNAs is faster or comparable to that for TFs, depending on the choice of kinetic parameters. For sRNA-based regulation, $\alpha_m = 3.5$ and $\alpha_s$ goes from $\approx 0.35$ to $4.5$ for switching from low to high and vice versa for high to low. For TF-based switching $\alpha_m$ is such that both schemes have same steady states. Other parameters as in Fig. \ref{fig:noiseburst}. }
\label{largesignal}
\end{figure}

\newpage 

\section*{Tables}

\begin{table}[h]
\begin{center}
\includegraphics[width=1.0\textwidth]{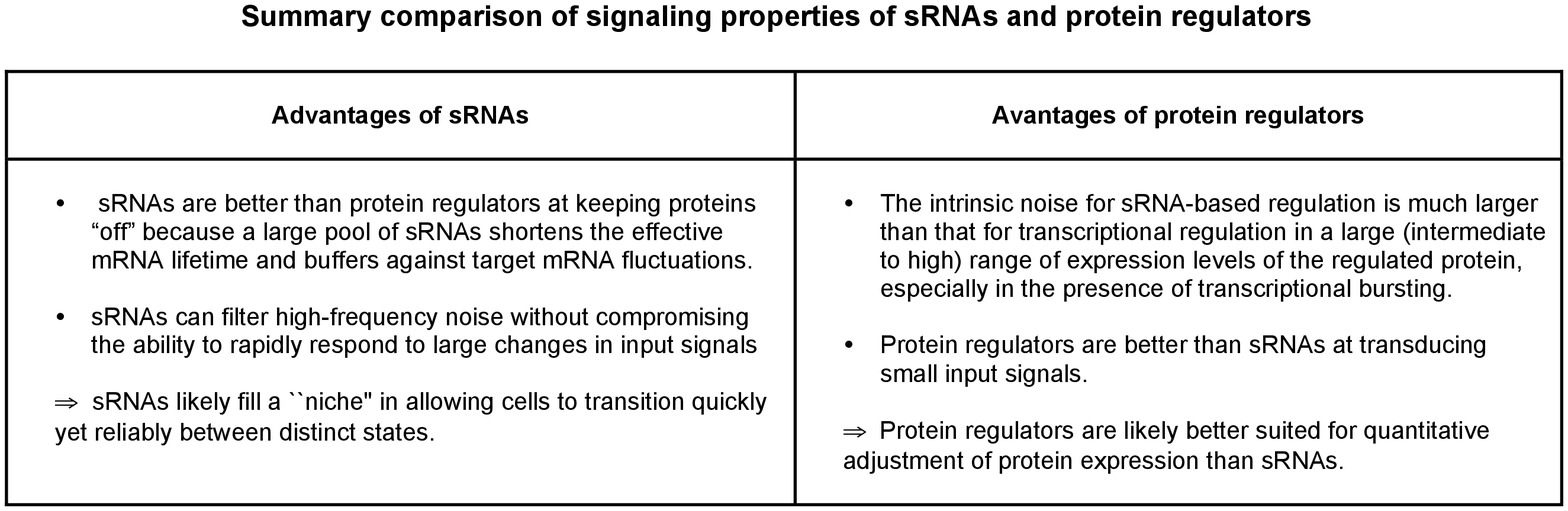}
\end{center}
\caption{ Table summarizing the advantages and disadvantages of sRNAs when compared to transcriptional protein regulators (transcription factors).}
\label{table}
\end{table}

\end{document}